\documentclass[a4paper,11pt]{article}
\usepackage{amsmath}
\usepackage{graphicx}
\usepackage{amsfonts}
\usepackage{amssymb}
\usepackage{subfigure}
\usepackage{tabularx}
\usepackage{sistyle}
\usepackage{feynmp}
\usepackage{indentfirst}

\newcommand{\psip}{\psi^{\prime}}
\newcommand{\psif}{J/\psi}
\newcommand{\etap}{\eta^{\prime}}

\begin{document}

\begin{center}
{\LARGE Ultimate survival in anomalous $\psi (2S)$ decays}\\
\vspace{0.5cm}
{\large{Jean-Marc G\'erard and Antony Martini}}
\\[12pt]{Centre for Cosmology, Particle Physics and Phenomenology
(CP3), }

\textit{Universit\'{e} catholique de Louvain, Chemin du Cyclotron 2, \\
 1348 Louvain-la-Neuve, BELGIUM}

\vspace{0.5cm}
\end{center}
\noindent\makebox[\linewidth]{\rule{16.9cm}{0.4pt}}
\textbf{{\large{Abstract}}} \\
\newline
The hierarchy among the radiative $\gamma$($\pi^0$, $\eta$, $\etap$) decay channels for the $\psip$ looks quite different from the $\psif$ one. The fate of charm, namely an ultimate survival of on-shell $c \bar{c}$ intermediate states, might give us the clue for this new puzzle in exclusive charmonium decays. A similar self-preservation has already been invoked in the past to solve the so-called $\rho \pi$ puzzle.\\
\noindent\makebox[\linewidth]{\rule{16.9cm}{0.4pt}}

\section*{Introduction}

Puzzling electromagnetic and strong anomalous two-body decays have been a source of successful inspiration in the past. In particular, the observation of non-vanishing $\pi^0 \rightarrow \gamma \gamma$ and suppressed $\Phi \rightarrow \rho \pi$ processes played a crucial role in our setting up of Quantum Chromodynamics (QCD). Interestingly, recent results on anomalous charmonium decays into light pseudoscalars \cite{Ablikim:2010dx} suggest a new puzzle. Indeed, the measured branching fractions given in Table \ref{PsiptoVPBR} imply rather different patterns for the $\psif$ and $\psip$ radiative modes respectively:

\begin{equation}
Br \left(\psif \rightarrow \gamma \pi^0 \right) \ll Br \left(\psif \rightarrow \gamma \eta \right) \simeq Br \left(\psif \rightarrow \gamma \etap \right)
\label{psifHierarchy}
\end{equation}

\begin{equation}
Br \left(\psip \rightarrow \gamma \pi^0 \right) \simeq Br \left(\psip \rightarrow \gamma \eta \right) \ll Br \left(\psip \rightarrow \gamma \etap \right)
\label{psipHierarchy}.
\end{equation}

So at the first glance $\eta$ appears to be on an equal footing with the other iso-singlet (i.e., $\etap$) in $\psif$ anomalous decays while, quite surprisingly, on equal terms with the neutral component of an iso-triplet (i.e., $\pi^0$) in $\psip$ ones. In this Letter, we consider the possibility that such a striking difference between the dynamics of the fundamental $c \bar{c}$ vector bound state and its first radial excitation is a direct consequence of the available phase space left for a $c \bar{c}$ meson to eventually survive $\psip$ anomalous decays.

\begin{table}[b]
\centering
\begin{tabularx}{16cm}{|c|>{\centering}X|>{\centering}X|>{\centering}X|}
\hline
 & $\gamma \pi^0 $ & $\gamma \eta $ & $ \gamma \etap $ \tabularnewline
\hline
$\psif$ & $\left(3.49 \pm 0.3 \right) \times 10^{-5}$ & $\left(1.104 \pm 0.034 \right) \times 10^{-3}$ & $\left(5.16 \pm 0.15 \right) \times 10^{-3}$ \tabularnewline
\hline
$\psip$ & $\left(1.6 \pm 0.4 \right) \times 10^{-6}$ & $\left(1.4 \pm 0.5 \right) \times 10^{-6}$ & $\left(1.23 \pm 0.06 \right) \times 10^{-4}$ \tabularnewline
\hline
\end{tabularx}
\caption{Experimental branching fractions for $\left[\left( \psif , \, \psip \right) \rightarrow \gamma \left( \pi^0 , \, \eta \,  , \, \etap \right) \right]$ \cite{PhysRevD.86.010001}.}
\label{PsiptoVPBR}
\end{table}

\section{Strong $c \bar{c}$ annihilation for $\psif \rightarrow \gamma P$} \label{Section1}
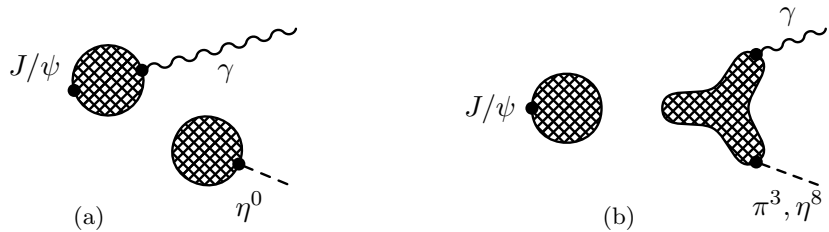
\begin{figure}[h]
  \label{StrongCCannihil}
  \centering
  \subfigure[]{\label{loop2AAhatched}
  \begin{fmffile}{loop2AAhatched}
  \begin{fmfgraph*}(180,60)
	\fmfleft{i}
	\fmfright{o1,o2}
	\fmf{phantom}{i,v1}
	\fmfv{label=$\psif$}{v1}
	\fmfpoly{smooth,tension=2.45,hatched}{v1,v2,v3,v4,v5,v6,v7,v8} 
	\fmf{phantom}{v3,v9}
	\fmfpoly{smooth,tension=0.45,hatched}{v9,v10}
	\fmf{photon,label=$\gamma$}{v5,o2}
	\fmf{dashes,label=$\eta^0$}{v10,o1}
	\fmfdot{v1,v5,v10}
  \end{fmfgraph*}
  \end{fmffile}}
  \subfigure[]{\label{loop2ghatched}
  \begin{fmffile}{loop2ghatched}
  \begin{fmfgraph*}(180,60)
	\fmfleft{i}
	\fmfright{o1,o2}
	\fmf{phantom}{i,v1}
	\fmfv{label=$\psif$}{v1}
	\fmfpoly{smooth,hatched}{v1,v2}
	\fmf{phantom}{v2,v3}
	\fmffixed{(0.8cm,0.0cm)}{v2,v3}
	\fmfpoly{smooth,pull=0.75,tension=0.45,hatched}{v3,v4,v5}
	\fmf{dashes,label=$\pi^3 ,, \eta^8$,l.side=right}{v4,o1}
	\fmf{photon,label=$\gamma$}{v5,o2}
	\fmfdot{v1,v4,v5}
  \end{fmfgraph*}
  \end{fmffile}}
\caption{Strong $c \bar{c}$ annihilation for \subref{loop2AAhatched} $\left[ \psif \rightarrow \gamma \eta^0 \right]$ and \subref{loop2ghatched} $\left[ \psif \rightarrow \gamma (\pi^3 \, , \, \eta^8) \right]$; multi-gluon exchanges between bubbles are not drawn.}
\end{figure}

Anomalous radiative $\psif$ decays into light ($q \bar{q}$) mesons are expected to proceed through the disconnected diagrams displayed in Fig.\ref{StrongCCannihil}.

Within a (questionable) perturbative QCD language, the two-bubble topologies in Fig.\ref{loop2AAhatched} and Fig.\ref{loop2ghatched} would simply amount to a $c \bar{c}$ annihilation into two and three gluons respectively. However, in the limit where topologies with a popped $s \bar{s}$ quark pair are identical to those with a popped $u \bar{u}$ or $d \bar{d}$, the full strong interaction dynamics behind Fig.\ref{StrongCCannihil} can be encoded in the following effective Lagrangian
\begin{equation}
{\cal{L}}_{strong} \left( \psif {\text{-}} \gamma {\text{-}} P \right) = \frac{2}{3} \, c_2 \, Tr(P) + c_3 \,  Tr(QP) \label{StrongLag}
\end{equation}
with $P=\pi^i \lambda_i$ the light pseudoscalar nonet and $Q=diag \left( \frac23, \, -\frac13, \, -\frac13\right)$ the quark electric charge matrix. In the isospin limit, the only $\pi$ mesons observed in the final states considered here are defined by 
\begin{align}
\pi^0 & \equiv \pi^3 = \frac{1}{\sqrt{2}} \left(u \bar{u} - d \bar{d}\right) \nonumber\\
\eta & \equiv \cos{(\theta)} \eta^8 - \sin{(\theta)} \eta^0 \stackrel{\theta_{ph}}{=} \frac{1}{\sqrt{3}} \left(u \bar{u} + d \bar{d} - s \bar{s} \right) \label{equation4}\\
\etap & \equiv \sin{(\theta)} \eta^8 + \cos{(\theta)} \eta^0 \stackrel{\theta_{ph}}{=} \frac{1}{\sqrt{6}} \left(u \bar{u} + d \bar{d} + 2s \bar{s} \right) \nonumber
\end{align}
where, for later references, the amount of strange/non-strange quarks in the wave functions is displayed for the so-called \cite{Degrande} phenomenological mixing angle $\theta = \theta_{ph} \cong \ang{-19.5}$ (or $\tan{(\theta_{ph})} = -{1}/{2 \sqrt{2}}$). This peculiar value for the $\eta^0 \text{-} \, \eta^8$ mixing angle is also compatible \cite{Gerard:2004gx} with the large $N_c$ limit expectation in QCD, $N_c$ being the number of colors.

The Lorentz-invariant structure for the anomalous decay amplitude of a vector meson $V$ into a photon $A$ and a pseudoscalar $\pi$ being given by $\epsilon_{\mu \nu \alpha \beta} \left\{ \partial^{\mu} V^{\nu} \partial^{\alpha} A^{\beta} \right\} \pi$, the resulting partial decay width is proportional to the cube of the $\pi$ momentum and the following ratios of branching fractions are easily inferred:

\begin{subequations}\label{equation5}
\begin{align}
 \left. \frac{Br \left( \psif \rightarrow \gamma \pi^0 \right)}{Br \left( \psif \rightarrow \gamma \eta \right)}\right|_{strong} &= 3 \left[ \frac{\sin{(\xi)}}{\sin{(-\theta + \xi)}} \right]^2 \frac{\left| \vec{p}_{\pi} \right|^3}{\left| \vec{p}_{\eta} \right|^3} \label{equation5a}\\
\left. \frac{Br \left( \psif \rightarrow \gamma \eta \right)}{Br \left( \psif \rightarrow \gamma \etap \right)} \right|_{strong} &= \tan^2{(-\theta + \xi)} \frac{\left| \vec{p}_{\eta} \right|^3}{\left| \vec{p}_{\etap} \right|^3} \label{equation5b}
\end{align}
\end{subequations}
with $\tan{(\xi)} \equiv -\left({c_3}/{c_2} \right)\tan{(\theta_{ph})}$. Once plugged into Eqs.(\ref{equation5}), the measured branching fractions listed in Table \ref{PsiptoVPBR} favor the values
\begin{equation}
\theta \cong \ang{-20.5} \; \mathrm{and} \; \xi \cong \ang{2.2}, \label{equation6}
\end{equation}
namely a $\theta$ mixing angle remarkably close to its phenomenological value $\theta_{ph}$ introduced in Eq.(\ref{equation4}) and a ratio ${c_3}/{c_2} = {\cal{O}}(1/10)$ quite compatible with a naive perturbative estimate based on the two ($c_2$) versus three ($c_3$) gluon annihilation processes understood in Fig.\ref{StrongCCannihil}. The inclusion in the strong Lagrangian (\ref{StrongLag}) of an $SU(3)$-breaking term proportional to the pseudoscalar matrix $P_{ss}$ (to weight the popped $s \bar{s}$ pair) may significantly affect the extracted value for the $\theta$ mixing angle. But in any case, the pattern (\ref{psifHierarchy}) for $\psif$ seems to be well explained in terms of strong annihilation decay amplitudes alone. Such is apparently not the case for the pattern (\ref{psipHierarchy}) as the relations (\ref{StrongLag}) and (\ref{equation5}) blindly applied to the corresponding $\psip$ data would then imply $\theta \cong \ang{-2.3}$ and $\xi^{\prime} \cong \ang{3.4}$, in flat contradiction with both phenomenological and perturbative QCD expectations $\theta < \ang{-10}$ and $\xi^{\prime} < \xi$, respectively. In other words, the puzzle raised when confronting (\ref{psipHierarchy}) with (\ref{psifHierarchy}) seems to lie fully upon the $\psip$. Actually the missing piece arises from $c \bar{c}$ survival for $\psip$, a particular feature which does not affect the fundamental $\psif$ state.

\section{Strong $c \bar{c}$ survival (or self-preservation) for anomalous $\psip$ decays}
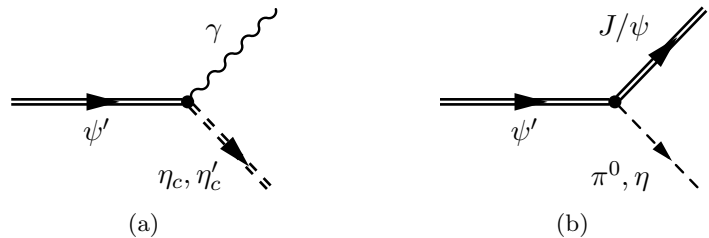
\begin{figure}[h]
  \centering
  \subfigure[]{\label{loop2Gg}
  \begin{fmffile}{loop2Gg}
  \begin{fmfgraph*}(110,70)
	\fmfleft{i}
	\fmfright{o1,o2}
	\fmf{heavy,label=$\psip$}{i,v1}
	\fmfdot{v1}
	\fmf{photon,label=$\gamma$,l.side=left}{v1,o2}
	\fmf{dbl_dashes_arrow,label=$\eta_c ,, \eta_c^{\prime}$,l.side=right}{v1,o1}
	\fmfset{arrow_len}{3mm}
  \end{fmfgraph*}
  \end{fmffile}} \hspace{1cm}
  \subfigure[]{\label{loop2AA}
  \begin{fmffile}{loop2AA-2}
  \begin{fmfgraph*}(110,70)
	\fmfleft{i}
	\fmfright{o1,o2}
	\fmf{heavy,label=$\psip$}{i,v1}
	\fmffixed{(2.3cm,0cm)}{i,v1}
	\fmfdot{v1}
	\fmf{heavy,label=$\psif$,l.side=left}{v1,o2}
	\fmf{scalar,label=$\pi^0 ,, \eta$,l.side=right}{v1,o1}
	\fmfset{arrow_len}{3mm}
  \end{fmfgraph*}
  \end{fmffile}}
\caption{Recorded survival in \subref{loop2Gg} $\left[ \psip \rightarrow \gamma (\eta_c, \eta_c^{\prime}) \right]$ and \subref{loop2AA} $\left[ \psip \rightarrow \psif (\pi^0, \eta) \right]$; the $\left[ \psip \rightarrow \psif \etap \right]$ mode is forbidden by kinematics.}
\label{topologiesFig}
\end{figure}

The observed anomalous $\psip$ decays into $c \bar{c}$ survivors are displayed in Fig.\ref{topologiesFig}.
These processes turn out to have relatively large branching fractions in spite of huge phase space suppressions. This empirical fact suggests the following rule: direct annihilations (analogous to the ones displayed in Fig.\ref{StrongCCannihil} for $\psif$) are negligible whenever an on-shell intermediate $c \bar{c}$ state is kinematically allowed. In other words, when possible, survival would precede any strong $c \bar{c}$ annihilation. And in the limit of an ultimate $c \bar{c}$ survival, the $\gamma(\eta_c,\eta_c^{\prime})$ together with the $\psif (\pi^0,\eta)$ two-body final states for $\psip$ decays would prohibit the $\gamma (\pi^0,\eta)$ ones. If such turns out to be the case, only $\gamma \etap$ can be produced through the (two-)gluon annihilation clearly at work in $\psif$ decays and the following simple pattern
\begin{equation}
Br \left(\psip \rightarrow \gamma \pi^0 \right) = Br \left( \psip \rightarrow \gamma \eta \right) = 0 \neq \left. Br \left( \psip \rightarrow \gamma \etap \right)\right|_{strong} \label{SimplePattern}
\end{equation}
emerges in a way from the very peculiar pseudo-Goldstone mass spectrum $m_{\pi , \eta}^2 \ll m_{\etap}^2$ of QCD.

Having at our disposal a qualitative understanding of the observed hierarchy (\ref{psipHierarchy}) from strong dynamics, let us now turn to the electromagnetic interactions to get more realistic $\psip \rightarrow \gamma \left( \pi^0 , \eta \right)$ branching fractions.

\section{Electromagnetic annihilation (or extinction) for $\psip \rightarrow \gamma P$}

\begin{figure}[t]
  \centering
  \begin{fmffile}{loop2EM}
  \begin{fmfgraph*}(180,60)
	\fmfleft{i}
	\fmfright{o1,o2}
	\fmf{phantom}{i,v1}
	\fmfv{label=$\psip$}{v1}
	\fmfpoly{smooth,hatched}{v1,v2}
	\fmf{photon,label=$\gamma$}{v2,v3}
	\fmffixed{(0.8cm,0.0cm)}{v2,v3}
	\fmfpoly{smooth,pull=0.75,tension=0.45,hatched}{v3,v4,v5}
	\fmfdot{v1,v4,v5}
	\fmf{dashes,label=$\pi^0 ,, \eta ,, \etap $,l.side=right}{v4,o1}
	\fmf{photon,label=$\gamma$}{v5,o2}
  \end{fmfgraph*}
  \end{fmffile}
\vspace{0.5cm}
\caption{Electromagnetic $c \bar{c}$ annihilation for $\left[ \psip \rightarrow \gamma P \right]$}
\label{ElectroAnnihil}
\end{figure}
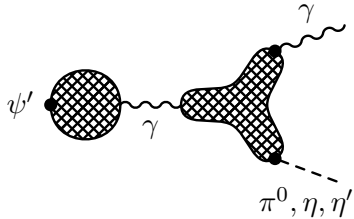

According to our survival hypothesis, the anomalous $\psip$ decays into $\gamma \pi^0$ and $\gamma \eta$ can only proceed through the diagram displayed in Fig.\ref{ElectroAnnihil}. These processes are encoded in the following effective Lagrangian
\begin{align}
{\cal{L}}_{em} \left( \psip \text{-} \gamma \text{-} P \right) &= \frac{2}{3} \, c_1^{\prime} \, Tr(Q^2 P) \nonumber \\
                                              &= \frac29 c_1^{\prime} \left\{ \frac23 Tr(P) +  Tr(QP) \right\} . \label{EMLag}
\end{align}

Comparing (\ref{EMLag}) with the effective Lagrangian (\ref{StrongLag}), we conclude that the electromagnetic $\psip \rightarrow \gamma P$ ratios of branching fractions due to Fig.\ref{ElectroAnnihil} are most easily derived from the strong $\psif \rightarrow \gamma P$ ones due to Fig.\ref{StrongCCannihil} by substituting $-\theta_{ph}$ for $\xi$ in the equations (\ref{equation5}). Doing so, one obtains:
\begin{subequations}\label{equation9}
\begin{align}
\left. \frac{Br \left( \psip \rightarrow \gamma \pi^0 \right)}{Br \left( \psip \rightarrow \gamma \eta \right)} \right|_{em} &= \frac13 \left[ \frac{1}{\sin{(\theta + \theta_{ph})}} \right]^2 \frac{\left| \vec{p}_{\pi} \right|^3}{\left| \vec{p}_{\eta} \right|^3}
\label{equation9a} \\
\left. \frac{Br \left( \psip \rightarrow \gamma \eta \right)}{Br \left( \psip \rightarrow \gamma \etap \right)} \right|_{em} &= \tan^2 {\left( \theta + \theta_{ph} \right)} \frac{\left| \vec{p}_{\eta} \right|^3}{\left| \vec{p}_{\etap} \right|^3} \; .
\label{equation9b}
\end{align}
\end{subequations}
On the one hand, the equation (\ref{equation9b}) confirms that a strong $c \bar{c}$ annihilation analogous to Fig.\ref{loop2AAhatched} is needed to enhance the $\gamma \etap$ channel with respect to the $\gamma \eta$ one since the electromagnetic $c \bar{c}$ annihilation alone in Fig.\ref{ElectroAnnihil} would require too small a $\tan^2 \left( \theta + \theta_{ph} \right)$ factor (namely, a positive $\eta^0 \text{-} \, \eta^8$ mixing angle). On the other hand, the relation (\ref{equation9a}) gives further support to our survival hypothesis since it implies a $\theta$-independent lower bound of $1/3$ for the $\gamma \pi^0$ branching fraction with respect to the $\gamma \eta$ one. Moreover, it leads to a definite range around unity,
\begin{equation}
0.8 < \frac{Br \left( \psip \rightarrow \gamma \pi^0 \right)}{Br \left( \psip \rightarrow \gamma \eta \right)} < 1.2 \; ,
\label{equation10}
\end{equation}
for realistic values of the $\eta^0 \text{-} \, \eta^8$ mixing angle (say, $\ang{-22} < \theta < \ang{-14}$) compatible with the one previously extracted from $\psif \rightarrow \gamma P$ in an $SU(3)$ limit (see Eq.(\ref{equation6})). Our key-prediction (\ref{equation10}) is in full agreement with the experimental value of $1.14 \pm 0.50$ derived from Table \ref{PsiptoVPBR}. Needless to emphasize that a more precise measurement of the $\psip \rightarrow \gamma \pi^0$ and $\psip \rightarrow \gamma \eta$ anomalous decays would provide us with a crucial test of the survival hypothesis defended here.

\section*{Conclusion}

In this Letter, we have argued that rather different hierarchies among anomalous charmonium decays are closely linked to the fate of charm and can be explained on the sole basis of an instinct of self-preservation. Looking for a dynamical explanation within QCD, at present we can only draw the following encouraging parallel between this survival hypothesis and the so-called OZI rule \cite{Okubo:1963fa}\cite{Zweig:1964jf}\cite{Iizuka:1966fk}. Originally invoked to explain the unexpected suppression of the $\rho \pi$ hadronic channel with respect to the $K \bar{K}$ one in $\Phi$ decays \cite{ZweigProc}, the OZI empirical rule that is now well understood within the inspiring large $N_c$ limit of QCD \cite{Hooft1974461} legitimates somehow our starting point for an ultimate survival in $\psip$ anomalous radiative decays, e.g.,
\[Br \left( \psip \rightarrow \gamma \eta_c, \psif \eta \right) \gg Br \left( \psip \rightarrow \gamma \eta \right) .\]\\
As a matter of fact, when the number of colors tends to infinity, the surviving $\eta_c$ does not mix with the light $q \bar{q}$ pseudoscalars, and the surviving $\psif$ has a narrow width.

At this level, it is also worth recalling that a softer survival scenario also implying no strong annihilation amplitude for the hadronic $\psip \rightarrow \rho \pi$ decay has already been invoked in the past \cite{Gerard:1999uf}\cite{Artoisenet:2005sg} to solve yet another $\rho \pi$ puzzle in anomalous $\psip$ decays \cite{Mo:2006cy}, namely the strong hierarchy
\[Br \left( \psip \rightarrow \rho \pi \right) \ll \frac{Br \left( \psip \rightarrow l^+ l^- \right)}{Br \left( \psif \rightarrow l^+ l^- \right)} \cdot Br \left( \psif \rightarrow \rho \pi \right) \] \\
violating the so-called $12 \%$ rule. The recent observations of $\psip \rightarrow \gamma P$ listed in Table \ref{PsiptoVPBR} might thus shed some new light on this early proposal. Indeed, strong $c \bar{c}$ survival in $\psip \rightarrow \psif \pi$ implies electromagnetic $c \bar{c}$ extinction for $\psip \rightarrow \rho \pi$, a request not incompatible with available data, as shown in \cite{Artoisenet:2005sg}\cite{Zhao:2006gw}.

\section*{Acknowledgements}

We thank Philippe Mertens for his careful reading of our manuscript. This work has been supported in part by the Belgian IAP Program BELSPO P7/37.

\bibliographystyle{arxivStyle}
\bibliography{FinalForArxiv_BIB}

\end{document}